\documentclass[preprint,12pt]{elsarticle}

\usepackage[left=30mm,right=25mm,top=30mm,bottom=25mm]{geometry}

\usepackage[utf8]{inputenc}
\usepackage{pgfplots,tikz,circuitikz}
\DeclareUnicodeCharacter{2212}{−}
\usepgfplotslibrary{groupplots,dateplot}
\usetikzlibrary{patterns,shapes.arrows}
\pgfplotsset{compat=newest}

\usepackage{lineno,hyperref}
\usepackage{amsthm,amssymb,amsmath,booktabs}
\usepackage{mathrsfs,epsfig,psfrag,fancyhdr,graphicx,graphics,lscape}
\usepackage{ae,float,multicol,mathtools}
\usepackage{gensymb,supertabular,graphicx,bbm,textcomp}
\usepackage[color,final]{showkeys}
\usepackage{balance}

\theoremstyle{plain}

\theoremstyle{definition}

\theoremstyle{remark}
\newtheorem{rem}{Remark}

\usepackage{amsmath,booktabs}
\usepackage{mathrsfs,epsfig,fancyhdr,graphics,lscape}
\usepackage{ae,float,multicol}
\usepackage{supertabular,bbm}
\usepackage{amssymb,fancyhdr,epsfig,psfrag}
\usepackage[color,final]{showkeys}
\usepackage{graphicx,eucal,mathrsfs,dsfont,rotating}
\usepackage{caption}
\usepackage{subcaption}

\usepackage{framed} 
\usepackage{multicol} 

\usepackage{nomencl} 
\makenomenclature
\renewcommand*\nompreamble{\begin{multicols}{2}}
\renewcommand*\nompostamble{\end{multicols}}

\def\1{{1\kern-.3468em{ 1}}}

\def\1{{1\kern-.3468em{ 1}}}

\def\stck[#1]#2{\stackrel{*}{#2} {\kern-0.6568em{{~}^{#1}}}}

\newcommand{\bref}[2][*]{\ifthenelse{\equal{#1}{*}}{(\ref{#2})} {(\ref{#1}\ensuremath{\text{--}}\ref{#2})}}

\usepackage{colortbl}
\usepackage{xcolor}

\colorlet{tableheadcolor}{gray!25} 
\newcommand{\headcol}{\rowcolor{tableheadcolor}} %
\colorlet{tablerowcolor}{gray!10} 
\newcommand{\rowcol}{\rowcolor{tablerowcolor}} %
\newcommand{\topline}{\arrayrulecolor{black}\specialrule{0.1em}{\abovetopsep}{0pt}%
            \arrayrulecolor{tableheadcolor}\specialrule{\belowrulesep}{0pt}{0pt}%
            \arrayrulecolor{black}}
\newcommand{\midline}{\arrayrulecolor{tableheadcolor}\specialrule{\aboverulesep}{0pt}{0pt}%
            \arrayrulecolor{black}\specialrule{\lightrulewidth}{0pt}{0pt}%
            \arrayrulecolor{white}\specialrule{\belowrulesep}{0pt}{0pt}%
            \arrayrulecolor{black}}

\newcommand{\bottomlinec}{\arrayrulecolor{tablerowcolor}\specialrule{\aboverulesep}{0pt}{0pt}%
            \arrayrulecolor{black}\specialrule{\heavyrulewidth}{0pt}{\belowbottomsep}}%

\usepackage{amssymb}

\usepackage{listings}

\definecolor{codegreen}{rgb}{0,0.6,0}
\definecolor{codegray}{rgb}{0.5,0.5,0.5}
\definecolor{codepurple}{rgb}{0.58,0,0.82}
\definecolor{backcolour}{rgb}{0.95,0.95,0.92}

\lstdefinestyle{mystyle}{
    backgroundcolor=\color{backcolour},   
    commentstyle=\color{codegreen},
    keywordstyle=\color{magenta},
    numberstyle=\tiny\color{codegray},
    stringstyle=\color{codepurple},
    basicstyle=\ttfamily\footnotesize,
    breakatwhitespace=false,         
    breaklines=true,                 
    captionpos=b,                    
    keepspaces=true,                 
    numbers=left,                    
    numbersep=5pt,                  
    showspaces=false,                
    showstringspaces=false,
    showtabs=false,                  
    tabsize=2
}

\begin{document}

\begin{frontmatter}
 \title{\LARGE
 Low-cost sensors and circuits for plasma education: characterizing power and illuminance 
 \tnoteref{t1}}

\author[utfpr,ucberkeley]{Alessandro N. Vargas\corref{cor1}\fnref{fn1}}\ead{avargas@utfpr.edu.br} 
\author[ucberkeley]{Victor Miller}
\author[ucberkeley]{Ali Mesbah}
\author[bologna]{Gabriele Neretti}

\cortext[cor1]{Corresponding author}

\address[utfpr]{Universidade Tecnol\'ogica Federal do Paran\'a, UTFPR, \\
Av. Alberto Carazzai 1640, 86300-000 Cornelio Proc\'opio-PR, Brazil.}
\address[ucberkeley]{Department of Chemical and Biomolecular Engineering, \\
University of California, Berkeley, CA 94720, USA.}
\address[bologna]{Electric Electronic and Information Engineering Department, \\ 
University of Bologna, Viale Risorgimento 2,
40136 Bologna, Italy.}

\tnotetext[t1]{Research supported in part
 by the Brazilian agency CNPq Grant 305158/2017-1; 305998/2020-0; 421486/2016-3.}

\begin{abstract}
Industrial applications of plasma have significantly increased beyond semiconductor manufacturing in recent years. This necessitates training a skilled workforce in plasma science and technology. However, an essential challenge to this end stems from the high cost of plasma devices and diagnostics.  The limited access to plasma devices has hindered plasma education, particularly in the least developed countries. To this end, this paper demonstrates how low-cost sensors and circuits can be developed to enable inexpensive plasma experiments in laboratory environments. In particular, we show how to measure high voltage, current, and power from a cold-atmospheric plasma discharge. Additionally, we develop a low-cost illuminance sensor and demonstrate how it can be used to estimate the corresponding plasma power. The low-cost sensors and electronics presented in this paper can aid educators in characterizing plasma power versus plasma illuminance.
\end{abstract}

\end{frontmatter}


 \section{Introduction} \label{introduction}

Plasma research has received a lot of attention over the last few decades, owing to its increased applications in science and technology \cite{lieberman2005principles,ChaMin2022,Weltmann2019,GEORGE2021109702}. For example, cold atmospheric plasma has been the critical element in materials processing \cite{DURAN2001743}, toxin reduction in food \cite{NEUENFELDT2023101045},  decontamination and sterilization \cite{Bruggeman_2009},  degrading pharmaceutical pollutants \cite{Trifi2023}, regulating combustion engines \cite{Alrashidi2018}, and increasing agriculture productivity \cite{Simek2021}.

To cope with the increasing demand for plasma education, educational plasma laboratories have been established, aiming to foster plasma knowledge \cite{gekelman2020,Rasmussen_2022}. However, creating an educational plasma laboratory is expensive---the cost can exceed tens of thousands of US dollars \cite{Yau_2020}. Moreover, creating research plasma laboratories can go even further---their cost can easily exceed millions of US dollars \cite{gekelman2020,Rej1994}.
Consequently, many researchers and educators remain away from plasma experiments, especially those in the least developed countries. For those individuals, the only option is to resort to simulations \cite{Yau_2020}.

Even though simulations represent an important educational tool, experienced educators advocate that students learn much more---and score much higher---when learning through hands-on experiments \cite{Shome22258,CORTER20112054}.
According to them, simulations do not replace hands-on experiments. 
 
To perform plasma experiments, a high-voltage power supply is needed. 
Some researchers and educators have attempted to make it more accessible. For example, a study describes a low-cost, high-voltage source discharging energy into a cathode-ray tube \cite{Saraiva2012}. More recently, a study has shown a DC power supply that generated up to 3kV to create gliding arcs \cite{Kim2023}.
Another study has reported how to assemble a low-cost 7kV DC power supply, yet it was not specifically designed to ignite plasma   \cite{Mottet9472977}.

A study has documented a low-cost, high-voltage sinusoidal generator with a peak voltage of up to 20kV \cite{neretti2019}. Another study has shown that one can assemble a nanosecond high-voltage generator with a cost of around U\$  2500 \cite{Maresca2020}; however, this cost is unaffordable for many individuals. Another device able to generate high voltage is known as {\it Marx generator} \cite{Zhong9600453,Aranganadin2022,Ren2021}, yet it is unclear whether this electronic topology can result in low-cost products. Together, these studies emphasize the importance of creating affordable high-voltage power supplies. Note, though, that a power supply is just one of the items required in plasma experiments.

Another indispensable item in any plasma experiment is sensors for plasma diagnostics.  Plasma diagnostics are used to understand properties of plasma, such as in assessing the gas composition and temperature through the optical emission spectroscopy in plasma \cite{CLAVE2021106111,golda2020vacuum,BIRYUKOV201875}, and in monitoring the voltage and current in plasma discharges \cite{Viegas_2020,Gidon8685139}; see also \cite{lieberman2005principles,gekelman2020,boulos2023handbook,lu2019nonequilibrium} for further discussion about other types of sensors deployed in plasma experiments.

Not surprisingly, sensors used in plasma experiments are expensive. For an illustration, we can mention that a high-voltage probe---key element to monitor the dynamic behavior of plasma discharge---can cost several thousand US dollars (see Section \ref{sec::probe::discussion}). 
  
  The main contribution of this paper is to develop low-cost sensors and circuits,
  and demonstrate how these sensors enable measuring (i) high voltage and current, (ii) plasma power, and (iii) plasma brightness (measured in lux). 
  All circuits and schematics are provided. All circuits are assembled with low-cost parts.


This paper does not intend to show technological advances; instead,  it focuses on a detailed description of low-cost sensors for plasma experiments. Even though the circuits and sensors were invented many decades ago and are ubiquitous in the electronics industry  \cite{electronics11213446,Shi2019,franco2001,sinclair2000sensors}, we bring them together for the sake of low-cost plasma experiments.
 Namely, we detail how to combine low-cost sensors and circuits, how to assemble them, and how to apply them in plasma experiments.
We then demonstrate how the low-cost sensors can be used to characterize plasma power versus brightness. To the best of the authors' knowledge, such an experiment relating plasma power to brightness has not been done before.

\begin{figure}[!t]
	\centering
	\includegraphics[width=10cm]{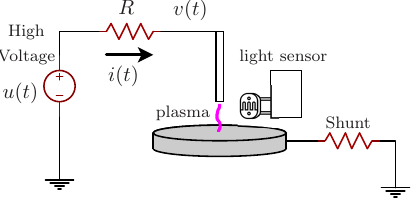}
	\caption{Setup of the plasma experiment. A high-voltage power supply provides direct current (DC) to the circuit. Plasma is ignited between a needle and a lower plate when the voltage $u(t)$ exceeds a certain threshold. A light sensor is conveniently placed to measure plasma brightness. A shunt resistor is used to help measure the current $i(t)$.}
	\label{fig::01}
\end{figure}

\section{Experimental setup}

This paper focuses on a direct current (DC) gliding arc discharge, a type of reactor that can generate cold atmospheric plasma \cite{Pei_2018,Cho2023,PEI2019217,Korolev_2014}. 
All the experiments detailed in this paper were performed in the open air and at atmospheric pressure, a usual configuration for local non-thermal treatments \cite{PEI2019217,Korolev_2014}.
Our setup allowed us to ignite plasma between a needle and a plate; see Fig. \ref{fig::01}. 

A high-voltage DC power supply generates an electric current $i(t)$ that flows through the circuit. Plasma is ignited when high voltage $u(t)$ is adjusted to exceed a certain threshold value. 
Note that $u(t)$ affects both the needle voltage $v(t)$ and the current $i(t)$. The power consumed by the plasma discharge then equals $p(t)=v(t)i(t)$ for all $t>t_0$, where $t_0$ denotes the time in which plasma is ignited. 
Conveniently placed next to the plasma discharge, 
a light sensor measures the plasma brightness $\ell(t)$, $t>t_0$. 

This paper shows how to measure $v(t), i(t),p(t)$, and $\ell(t)$, $t>t_0$. 

\begin{rem}
This paper does not cover how to build up a high-voltage direct current (DC) power supply. As mentioned previously, other studies have reported how to build high-voltage power supplies using both DC and AC. Implementing such devices is outside the scope of this paper.
\end{rem}

{\it Warning:} The experiments described in this paper were carried out with an industrial 
high-voltage DC power supply manufactured by Spellman model SL10PN1200 (Hauppauge, NY, USA), see Section \ref{sec::experiments}. This device complies with the US regulations for electrical safety and protection. Plasma experiments involve high voltage, which may even ensue the risk of death. High voltage can be lethal. For this reason, only trained, qualified individuals should attempt to deal with high voltage. All the low-cost sensors and electronics described in this paper manipulate high voltage. Only qualified personnel should attempt to reproduce the circuits shown in this paper. All the low-cost sensors and circuits were tested in a laboratory under safe working conditions. No electric problems were observed during the experiments.

\section{Low-cost sensors and circuits}\label{sec::sensors}

This section details how to implement the sensors and circuits necessary to measure the physical properties shown in Fig. \ref{fig::01}.

\subsection{Low-cost sensor for plasma brightness}\label{sec::lowcost::sensor}

Researchers of plasma science often rely on optical emission spectrometers to measure plasma light properties (e.g., \cite{Xingxing2017}, \cite{Nie2017}). For example, optical emission spectrometers can measure plasma intensity as a function of wavelength so that researchers can point out certain properties of plasma chemistry \cite{CLAVE2021106111,golda2020vacuum,BIRYUKOV201875}, \cite[Fig. 4.12, p. 62]{lu2019nonequilibrium}. Indeed, optical emission spectrometers are the go-to instruments for plasma emission analysis. But they are expensive. The starting cost of simple models ranges from thousands of US dollars. This means that optical emission spectrometers remain inaccessible to most researchers based in low-income countries and laboratories with meager financial support.


As an attempt to circumvent this affordability problem, we devised a low-cost circuitry for a light sensor. This sensor is limited because it does not measure the full scale of wavelenghts. However, as detailed in the sequel, it does measure the aggregate of wavelenghts in the form of {\it illuminance}. 

\subsection{High-voltage probe and current probe for measuring plasma power}\label{sec::probe::discussion}

Measuring voltage and current in plasma discharge is not a novel research topic. Researchers have been doing so for many decades;  see, for instance, the voltage-current curves in \cite[Fig. 14.2, p. 538]{lieberman2005principles},
\cite[Fig. 1]{Gudmundsson_2017}, and
\cite[Fig. 2]{Conrads_2000}. Voltage and current combined determine the plasma power, a key information for researchers.
For example, researchers monitor power because it changes the plasma properties when in contact with a target material skin  \cite{roy2022tolerance,TianDi2023} and in material processing \cite{SAMAL20173131,KASEEM2021100735}.

Researchers have almost exclusively documented the use of industrial probes to characterize the voltage-current relationship in plasma.
For instance, experiments were made with the probe model P6015A in
\cite{Viegas_2020,Gidon8685139,LiJing2020}, model P5210 in \cite{Pachuilo6877676}, model TCP312A in 
\cite{Dhakar9383175}, and 
model TCP0020 in \cite{Song2021},
all of the probes manufactured by
Tektronix (Beaverton, OR, USA). 

Industrial probes, like the aforementioned ones, are precise and extremely useful. However, they are expensive.
This paper demonstrates that it is possible to assemble reliable, low-cost probes, as described next.

\begin{figure}[!t]
	\centering
	\includegraphics[width=12cm]{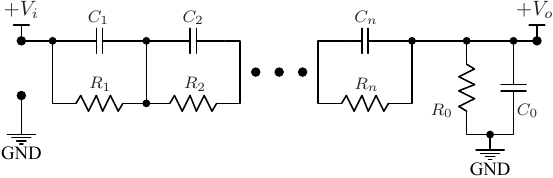}
	\caption{Schematics of the high-voltage probe. When high-voltage input is applied in $+V_i$, a low voltage appears in the output terminal $+V_o$. A series of  RC filters treats the corresponding signal and defines the attenuation $V_o/V_i$. The value of the components $C_0,\ldots,C_n$ and $R_0,\ldots,R_n$ varies according to the project specification.}
	\label{fig::02}
\end{figure}

\subsubsection{Low-cost high-voltage probe}\label{sec::lowcost::voltage::probe}

Studies have documented that a simple, low-cost resistive voltage divider can effectively measure high voltage \cite{electronics11213446,Shi2019}. As suggested by the authors of 
\cite{electronics11213446}, the scheme of Fig.  \ref{fig::02} suffices for
measuring high voltage. Measurements from the circuit of Fig. \ref{fig::02} can generate low-distortion signals for frequencies of up to 1 MHz (see Fig. 9 in \cite{electronics11213446}). This finding is confirmed in another study \cite{Shi2019}.  We followed these studies and implemented the circuit of Fig.  \ref{fig::02}.

{\it Implementation}:   
The circuit of Fig.  \ref{fig::02} was assembled in the laboratory with $C_1=\cdots=C_n$ and $R_1=\cdots=R_n$. This condition allows us to write (see Appendix)
\[
\frac{V_o}{V_i} = \frac{R_0}{n R_1+R_0}, \quad
\text{and} \qquad
C_0 = \frac{C_1R_1}{R_0},
\]
where $V_i$ represents the high-voltage input and $V_o$ denotes the low-voltage output.
Our low-cost probe has attenuation close to 1000:1. To obtain this attenuation, we set $n=5$, and picked up off-the-shelf components with values 
  $C_i= 15$pF, $R_i=10$M$\Omega$, $i=1\ldots,5$,
  $C_0 = 3$nF, and $R_0 = 52.8$K$\Omega$.
All resistors and capacitors were rated for high voltage for safety reasons (see further details in Appendix). 
  The assembled probe is depicted in Fig. \ref{fig::12}. Epoxy resin covers the components as these chemicals increase the insulation between the components' terminals, creating extra protection against arcs or accidental short circuits.

\begin{figure}[!t]
	\centering
	\includegraphics[width=9cm]{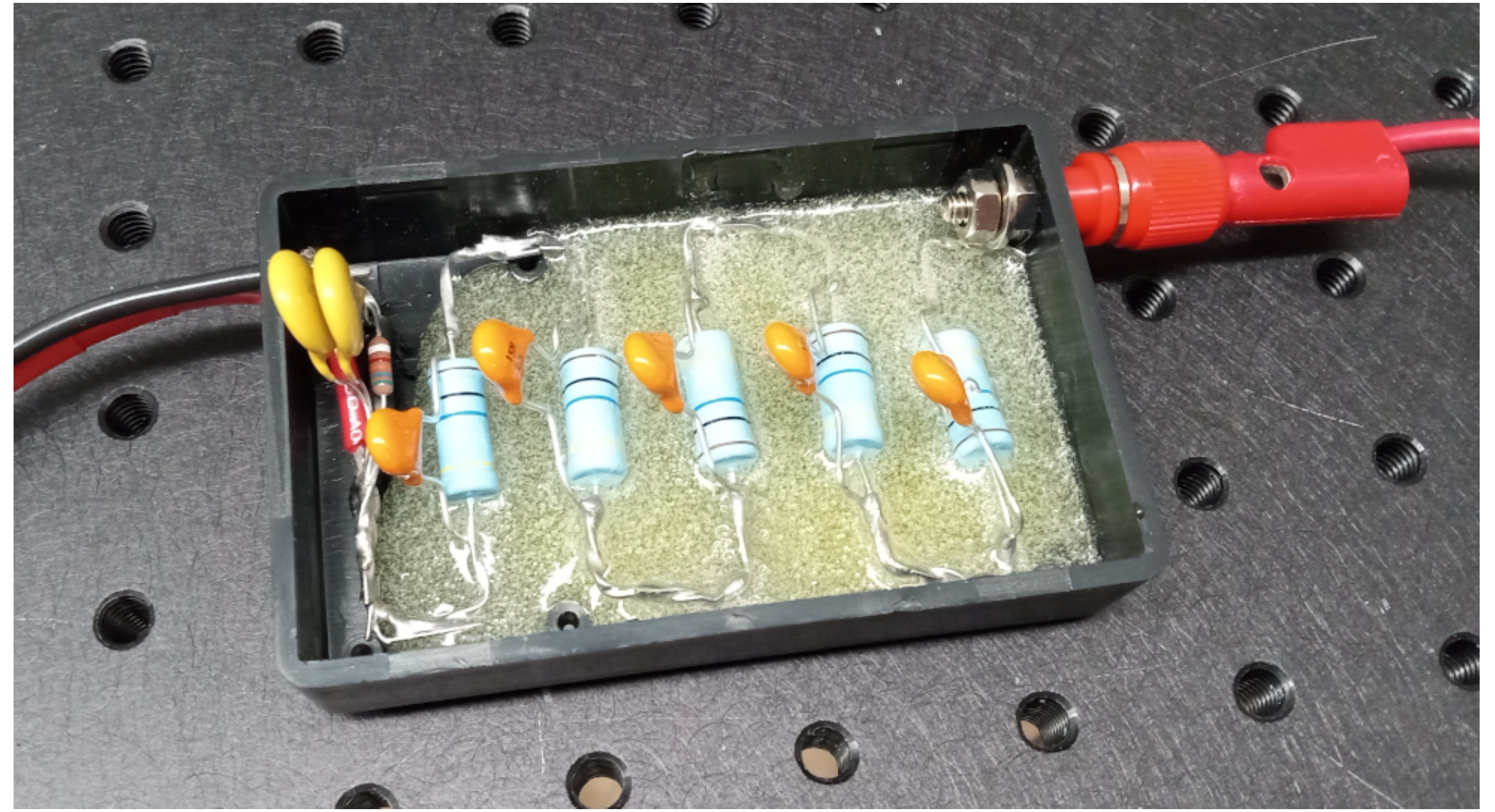}
	\caption{Low-cost high-voltage probe.
	}
	\label{fig::12}
\end{figure}

\begin{figure}[!t]
	\centering
	\includegraphics[width=7.5cm]{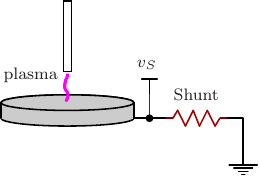}
	\caption{Configuration of the shunt resistor. This scheme allows us to measure the current that flows through the plasma discharge.
	}
	\label{fig::03}
\end{figure}

\subsubsection{Low-cost current probe}
\label{sec::lowcost::current::probe}

When a resistor is connected in series with a circuit, we immediately observe a voltage drop in the resistor's terminals. This is the working principle 
of the {\it shunt resistor} (e.g.,  \cite[p. 31]{Weranga2014}, \cite{Ziegler4797906}), a classical current sensor \cite[p. 31]{Weranga2014}.
A shunt resistor can measure both alternating current (AC) and direct current (DC).

The shunt resistor has been a popular solution to measure current in plasma experiments (e.g., \cite[Fig. 1]{SALEEM2020123031}, \cite[Fig. 1]{Gidon8685139}, \cite[Fig. 1]{GIDON2021104725}). As usual in the plasma literature, the shunt resistor connects the metal plate and the ground (Fig. \ref{fig::03}). The shunt resistor's value should be small enough to consume as little plasma power as possible.
 
\begin{figure}[!t]
	\centering
	\includegraphics[width=9cm]{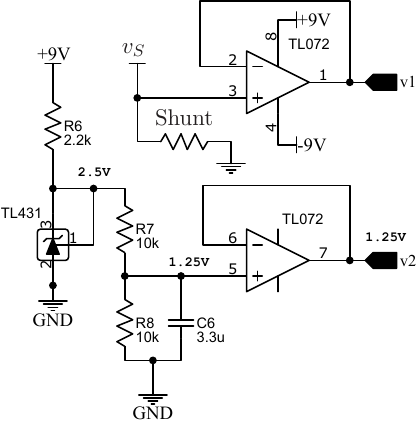}
	\caption{Auxiliary circuit for the current sensor. Upper circuit: buffer for the signal $v_S$ from a shunt resistor. Lower circuit: DC offset reference fixed at $1.25$V.
	}
	\label{fig::04}
\end{figure}

The circuit of Fig. \ref{fig::04} allows us to measure the plasma current.  The circuit contains a low-cost operational amplifier TL072 and a low-cost voltage reference TL431, both manufactured by Texas Instruments (Dallas, TX, USA). The upper part of Fig. \ref{fig::04}
shows a buffer stage that ensures high impedance for the signal $v_S$; the lower part generates a DC offset fixed at $+1.25$V.  The reason for generating this DC offset is related 
to a limitation on the microcontroller analog input, as detailed next.

Most microcontrollers have analog-to-digital converters that can handle only positive voltages. In this case, a microcontroller cannot read a negative voltage on the shunt voltage $v_S$. However,  we can add the negative signal $v_S$ to the DC offset of $1.25$V and send the corresponding signal to the microcontroller if $v_S+1.25$ is positive. This strategy allows us to consider negative voltages on $v_S$ up to $-1.25$V. 

\begin{figure}[!t]
	\centering
	\includegraphics[width=8.25cm]{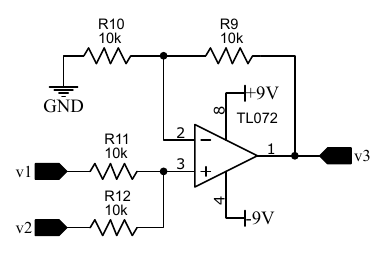}
	\caption{Non-inverting summing amplifier. The signal at the point v3 equals v1+v2.
	}
	\label{fig::13}
\end{figure}

{\it (Implementation)}.
The circuits of Figs. \ref{fig::03}--\ref{fig::13}  were implemented in the laboratory.
A shunt resistor of $23$ $\Omega$ was used (it was the only precise resistor available during the experiments). From Fig. \ref{fig::04}, we have v1$=v_S$ and v2=$1.25$V. Both v1 and v2 were connected in the corresponding ports of the circuit shown in Fig. \ref{fig::13}, which in turn guarantees that v3=v1+v2=$v_S+1.25$.

\subsubsection{Light-dependent resistor}
A light-dependent resistor (LDR) is a device that changes its resistance according to the light intensity (e.g., \cite[Sec. 14.4]{fraden2004}, \cite[Sec. 3.5]{sinclair2000sensors},\cite[Sec. 1.6.2]{haraoubia2018nonlinear}). LDR is a semiconductor device manufactured so that its two-terminal resistance decreases when the light intensity hitting its surface increases. The relationship between resistance and light intensity is nonlinear, as manufacturers indicate (e.g., \cite{Coramik2021}, \cite[p. 2]{roman2020light}).

This paper shows that LDR sensors are appropriate for plasma experiments. Indeed, the reasons for using LDR in plasma experiments are as follows. First, LDR is small---it has round faces under a few millimeters in size, taking up a small space in plasma reactors. Second, LDR is cheap---its price is less than three US dollars each. Third, LDR is easy to use and behaves like a classical resistor. These features have led to the widespread of LDR in industry, such as in the circuits that control the street lights \cite[Sec. 2.1.1]{thungtong2021web} and that measure radiotherapy irradiation  \cite{s20061568}.

As with any semiconductor-based component, LDR has no calibration. Namely, during the manufacturing process, each LDR sensor is made unique. As a result, each LDR inherits a particular light-sensing characteristic that can differ widely from other LDRs manufactured in the same industrial line. Two off-the-shelf LDR sensors---manufactured by the same company and sold under the same part number---might behave completely differently. 

Manufacturers openly disclose the unmatched properties of their LDRs in manuals; see, for example, how the resistance of an LDR can vary drastically from component to component, as indicated in the datasheet \cite[Fig. 1]{PDVP9008}. 
 Therefore, a fair use of LDR mandates a calibration. The next section presents a method to calibrate LDRs, borrowing ideas from \cite{Coramik2021}.

\begin{figure}[!t]
	\centering
	\includegraphics[width=8.25cm]{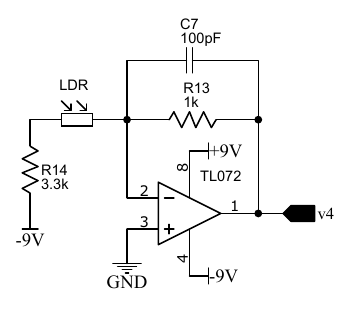}
	\caption{LDR amplifier circuit.
	}
	\label{fig::14}
\end{figure}

\subsection{Calibration of a light-dependent resistor}
  \label{sec::calibration}
 
The calibration experiment was performed with
the circuit of Fig. 
\ref{fig::14} (this circuit is borrowed from \cite[Fig. 2.3, p.63]{franco2001}). The LDR was manufactured by Advanced Photonics, photoresistor model PDV-P9008 (Palo Alto, CA, USA). 

\begin{figure}[!t]
	\centering
	\includegraphics[width=10cm]{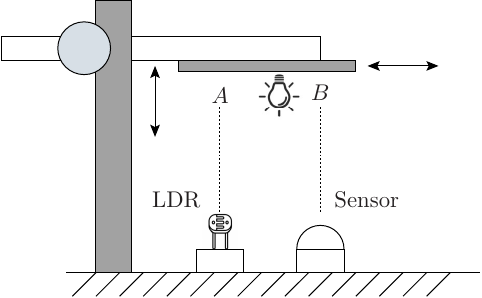}
	\caption{Setup for the calibration of a light-dependent resistor (LDR). The sensor on the right represents an industrial ready-to-use illuminance meter.}
	\label{fig::05}
\end{figure}

To calibrate this LDR, we followed the procedure suggested in \cite{Coramik2021}. The author of \cite{Coramik2021} suggests using a boom stand to hold the light source; see Fig. \ref{fig::05}.
Our boom stand had a holding arm that could move up and down.  A clamp was adjusted to set the arm's height when necessary. The arm was attached to a small sliding table that could move freely in the horizontal direction. Glued at the lower part of the sliding table was a small automotive light bulb manufactured by  
Sylvania model 1157A (Wilmington, MA, USA).

Our boom stand allowed us to position the light bulb at any point. In particular, we aligned the light bulb with points $A$ and $B$, corresponding to the perpendicular position of two sensors fixed at the ground: (A) LDR sensor (model PDV-P9008) and (B) industrial light meter sensor, manufactured by 
Urceri model MT-912 (China---sold at Amazon.com), which allowed us to measure illuminance (lux).

\begin{figure}[!t]
	\centering
	\includegraphics[width=11cm]{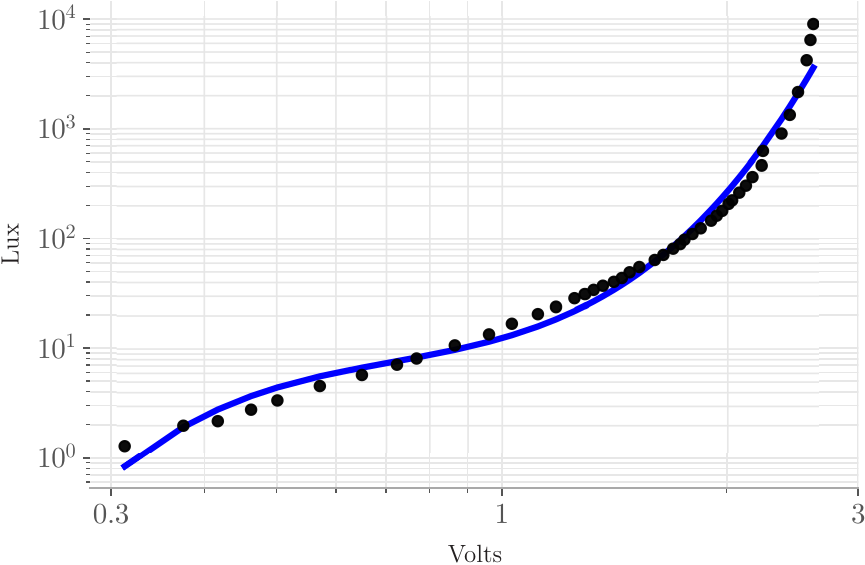}
	\caption{Experimental data from the LDR calibration in logarithm scale. The dots represent real data measured in the laboratory. The curve represents a third-order polynomial that fits the data.}
	\label{fig::06}
\end{figure}

All the calibration experiments were performed in a dark room. We applied values from 10V to 18V to the light bulb and moved it from point A to point B. 
The arm's height was decreased in steps until the light bulb reached a few millimeters away from the two sensors. 
The corresponding experimental data is summarized in Fig. \ref{fig::06}. 
As the data plotted in  Fig. \ref{fig::06} suggest, measurements from the LDR can be approximated by a third-order polynomial, which reads as 
\begin{equation} \label{eq::polynomial}
    \Delta(s) = a_3s^3 + a_2s^2+a_1s+a_0,
\end{equation}
where the values for $a_0,\ldots,a_3$ are available in Table \ref{tab::01} (see Appendix). 

Because \eqref{eq::polynomial} follows a logarithm scale, we now introduce its linear counterpart. Let $\ell$ denote the illuminance measurement from the LDR, and consider $v$=v4 as in Fig. \ref{fig::14}. Setting $\ln(\ell) = \Delta(s)$ and $\ln(v)=s$, we have from  \eqref{eq::polynomial}  that 
\[\ln(\ell) = a_3\ln(v)^3 + a_2\ln(v)^2+a_1\ln(v)+a_0,
\]
which is equivalent to
\begin{equation} \label{eq::polynomial::02}
    \ell = \exp\left(a_3\ln(v)^3 + a_2 \ln(v)^2+a_1\ln(v)+a_0 \right).
\end{equation}

As detailed in the next section, the formula in \eqref{eq::polynomial} was programmed in a microcontroller to convert voltage to illuminance. For illustration, suppose that the microcontroller reads +1V at the point v4. It then sets $v=1$ in the expression  \eqref{eq::polynomial} to obtain
$\ell = 12.5952$ (lux); see further details in Remark 
\ref{rem::free::documentation}.

\subsection{Measuring plasma power}\label{sec::plasma::power}

A microcontroller Arduino, model Due (Monza, Lombardia, Italy), was used to process experimental data. 
Arduino is a popular brand that shares free, open-hardware platforms, allowing users to modify its hardware and software as much as needed.
Arduino Due has twelve analog-to-digital converter (DAC) ports and two digital-to-analog converters (ADC). The DAC ports can handle voltage from 0V to 3.3V; the ADC ports can generate voltage from 0.55V to 2.75V.  
Arduino Due belongs to the family of low-cost microcontrollers (see \url{www.arduino.cc}).

In the laboratory, the low-cost high-voltage probe measured the voltage $v(t)$ at the needle
(see Section \ref{sec::lowcost::voltage::probe});  and the low-cost current probe measured the current $i(t)$ flowing through the plasma (see  Section \ref{sec::lowcost::current::probe}). 
Information from both probes feeds the two DAC ports of the Arduino Due, which then computes the plasma power $p(t) = v(t)i(t)$, for all $t>t_0$. 

\begin{figure}[!t]
	\centering
	\includegraphics[width=12cm]{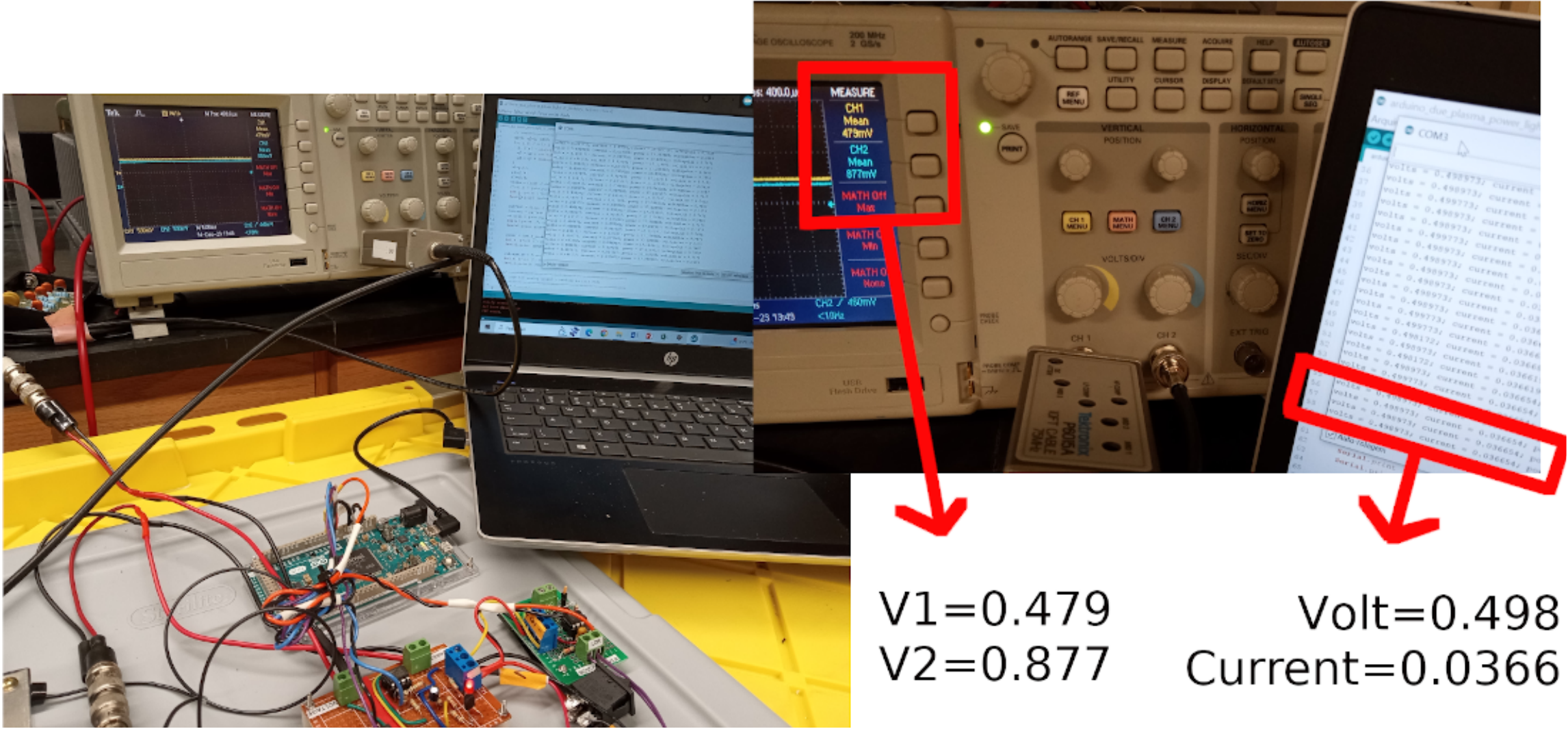}
	\caption{Low-cost sensors and electronics for 
 plasma experiments. Sensors measured real-time data from a plasma discharge and sent the data to Arduino Due. A laptop received data from Arduino Due and projected on-screen information about the voltage $v(t)$, current $i(t)$, power consumed by the plasma $p(t)$, and plasma brightness $\ell(t)$.}
	\label{fig::18}
\end{figure}

Arduino Due sends real-time data to a computer screen, allowing users to monitor the plasma power. For example, Fig. \ref{fig::18} shows a snapshot of measurements taken from both Arduino Due and an oscilloscope Tektronix model TDS2022B (high-voltage probe  Tektronix P6015A on Channel 1 and oscilloscope probe reading the voltage on shunt resistor on Channel 2).
As indicated in Fig. \ref{fig::18}, the consumed power was 
$p(t) = 0.498\times 10^3\times 0.0366 = 18.227$ (watts)
according to Arduino Due; and it was $p(t) = 0.479\times 10^3 \times (0.877/23) = 18.264$ (watts) according to the industrial probes. 
These numbers suggest that the power computed by our low-cost approach differs less than 2\% from the industrial instrument.


\begin{figure}[!t]
	\centering
	\includegraphics[width=11cm]{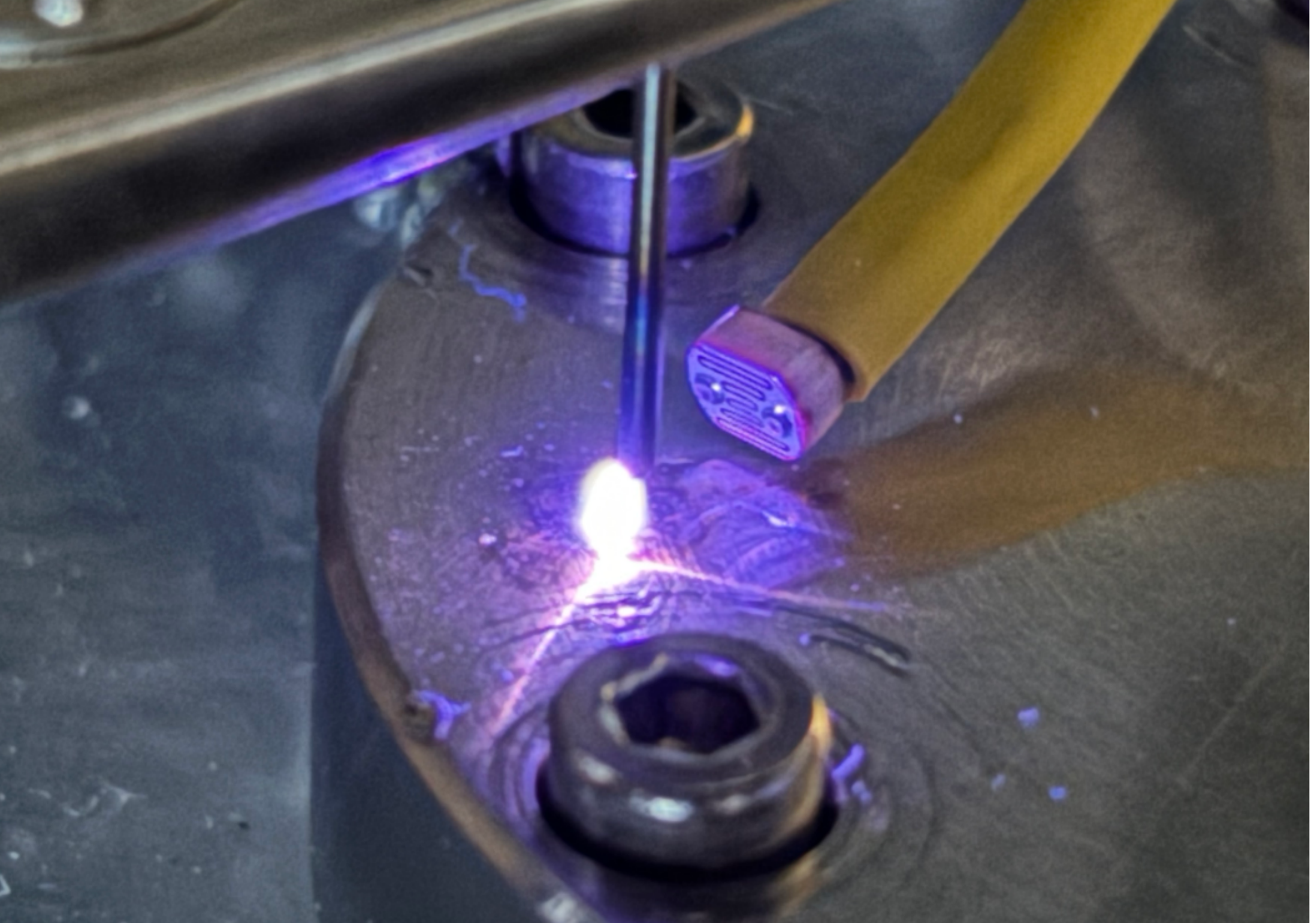}
	\caption{Plasma experiment. A light-dependent sensor (LDR) was positioned close to the plasma discharge to measure plasma brightness while the power consumed by the plasma was monitored.}
	\label{fig::16}
\end{figure}

\section{Experiments: characterizing plasma power versus illuminance}
\label{sec::experiments}

The plasma experiments were carried out under the scheme shown in Fig. \ref{fig::01}. A high-voltage direct current (DC) power supply, manufactured by  Spellman model SL10PN1200 (Hauppauge, NY, USA), was used to generate up to +10kV (100mA) in one terminal of a ballast resistor of $R=120$K$\Omega$.
This resistor limits the maximum discharge current to below 120mA, the maximum value the DC power supply can provide \cite[p. 219]{PEI2019217}. The other resistor's terminal was connected to a tungsten electrode needle (it contained 2\% lanthanide) with a diameter of 1.016 millimeters, manufactured by Diamong Ground Products Inc. (Newbury Park, CA, USA).

A stainless steel plate was placed below the electrode needle, with a distance of about five millimeters.
A light-dependent resistor (LDR), already described in Section \ref{sec::calibration}, was placed about eight millimeters away from the plasma discharge. 

The needle-to-plate configuration of Fig. \ref{fig::01} formed a plasma discharge (see Fig. \ref{fig::16}). 
While plasma was under production,  the needle-to-plate discharge was kept inside a chemical fume hood for safety. All the experiments were conducted in the open air, with the temperature at about $24^{\circ}$C and at atmospheric pressure.

 Arduino Due processed the voltage corresponding to the plasma brightness (i.e., illuminance) from the LDR (see Section \ref{sec::calibration}), following the expression \eqref{eq::polynomial}.  Arduino Due also processed the voltage $v(t)$ and 
 the current $i(t)$ (see  Section \ref{sec::plasma::power}) to compute the power plasma $p(t) = v(t)i(t)$, for all $t>t_0$.  
 
\begin{figure}[!t]
	\centering
	\includegraphics[width=11.5cm]{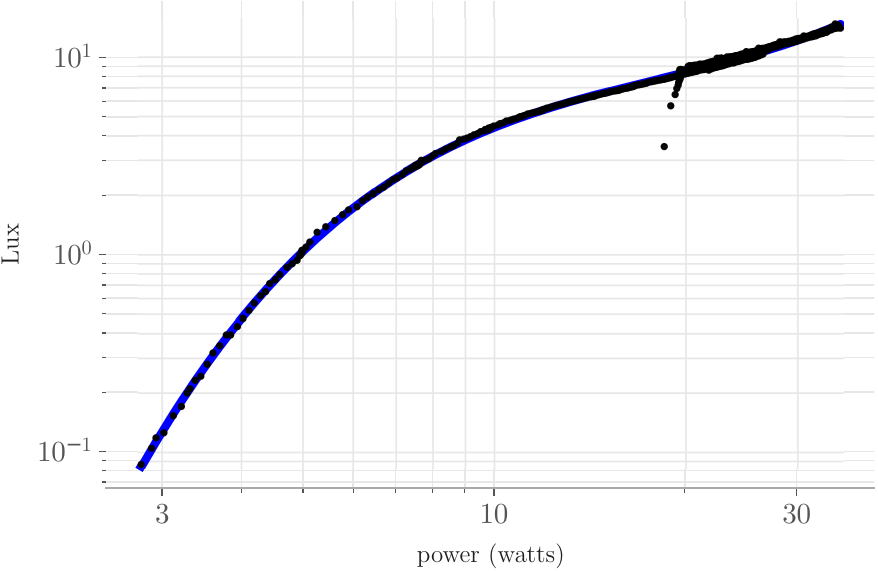}
	\caption{Experimental data from a plasma discharge. A third-order polynomial (blue) fits the experimental data (dot points).}
	\label{fig::08}
\end{figure}

The DC high-voltage power supply 
was manipulated to vary voltage between 0V and 10kV.
Plasma was ignited when the voltage crossed a threshold of about 4.5kV.
Once the plasma was ignited, it remained in production even under voltages moderately below the threshold. 
All the experimental data were recorded and are freely available for download; see Remark \ref{rem::free::documentation} in connection.

Fig. \ref{fig::08} summarizes the corresponding experimental data. It shows the relation between the power consumed by the plasma discharge and the plasma brightness.  The curve in Fig.  \ref{fig::08} is a third-order polynomial, approximated through experimental data (see Section \ref{sec::calibration})
\begin{equation} \label{eq::polynomial::03}
    \ell = \exp\left(a_3\ln(p)^3 + a_2 \ln(p)^2+a_1\ln(p)+a_0 \right),
\end{equation}
where $p$ represents the plasma power (watts), and $\ell$ denotes the plasma brightness (lux); the constants $a_0,\ldots,a_3$ are available   
in Table \ref{tab::01} (see Appendix). 

Experimental data suggest that the polynomial in \eqref{eq::polynomial::03} can represent plasma power versus plasma illuminance (Fig.  \ref{fig::08}). However, the polynomial does not capture the transient behavior of plasma dynamics. 
For example, Fig. \ref{fig::08} shows a minor glitch in the experimental data, corresponding to data recorded once the plasma was ignited. In other words, Arduino Due started recording the illuminance and power once the plasma ignited. Thus, in Fig.  \ref{fig::08}, the transient points correspond to the black dots located away from the blue curve.

In summary, the expression in 
\eqref{eq::polynomial::03} does characterize the relation between the plasma power and plasma brightness, yet it comes 
from measurements made with low-cost sensors and electronics. This evidence illustrates the potential of low-cost electronics for plasma experiments. Besides, researchers and educators can estimate the plasma power by measuring the plasma luminescence 
and vice versa, thus dispensing expensive industrial solutions.

Our initiative helps spread the knowledge of experimental plasma---in particular plasma power and plasma brightness---among laboratories with a limited budget, facilitating education on plasma.

 \begin{rem}\label{rem::free::documentation}
 All the files used in this manuscript---all files containing data, source code, schematics, and circuit layouts---are freely available on GitHub  at 
 \url{https://github.com/labcontrol-data/plasma-lowcost}
  Table \ref{tab::02} (Appendix) shows a final bill of materials used in this project.
 \end{rem}

\section{Concluding remarks}

This paper showed how to use low-cost sensors and electronics can be used to characterize plasma experiments. With a budget of less than U\$ 120, suitable circuits were assembled to measure high voltage, current, plasma power, and plasma illuminance. Such low-cost solutions can represent a paradigm shift in plasma education and research, particularly in under-resourced communities.



\section*{Appendix}

{\it Frequency evaluation of the high-voltage probe shown in Fig. \ref{fig::02}.} 
It is well-known that a resistor and a capacitor---when in parallel---create impedance equal to (e.g., \cite[p. 1526]{Domansky2019})
\[Z_i(s) = \frac{R_i}{1+R_iC_is}, \quad i=0,\ldots,n,
\]
where `$s$' denotes the complex variable in the Laplace domain (e.g., \cite[Ch. 3]{franklin2015feedback}). The input-output relation of the schematics of Fig. \ref{fig::02}  equals to
\begin{equation}\label{eq::laplace::01}
    G(s) = \frac{V_o(s)}{V_i(s)} = \frac{Z_0(s)}{\sum_{i=0}^n Z_i(s)}.
\end{equation}
Substituting $C_1=\ldots=C_n$ and $R_1=\ldots=R_n$ in \eqref{eq::laplace::01} yields
\begin{equation}\label{eq::laplace::02}
    G(s) = \frac{R_0(1+R_1C_1s)}{nR_1+R_0 + (nR_0R_1C_0+ R_0R_1C_1)s}.
\end{equation}
In \eqref{eq::laplace::02}, we substitute
\[C_0 = \frac{R_1C_1}{R_0},
\]
to obtain
\[
G(s) = \frac{R_0(1+R_1C_1s)}{nR_1+R_0 + (nR_1^2C_1+ R_0R_1C_1)s}.
\]
Because $G(s)$ equals to
\[
      \frac{(1+R_1C_1s)}{\frac{nR_1+R_0}{R_0} + \frac{(nR_1^2C_1+ R_0R_1C_1)}{R_0}s} 
      =  
      \frac{(1+R_1C_1s)}{\frac{nR_1+R_0}{R_0} (1+R_1C_1)s},
\]
we have that 
\begin{equation}\label{eq::laplace::03}
G(s) = \frac{R_0}{nR_1 + R_0}.    
\end{equation}
Note in \eqref{eq::laplace::03} that $G(s)$ does not depend on the variable $s$. As a result, the input-output gain of the probe remains constant for all possible frequencies, and the phase shift equals zero 
for all possible frequencies \cite[Ch. 6.1]{franklin2015feedback}.

\begin{table}[!t]
\centering
    \caption{Parameters representing two equations: (i) LDR sensor function as in \eqref{eq::polynomial}, and (ii) plasma power versus plasma brightness as in \eqref{eq::polynomial::03}. \label{tab::01}}
    \begin{tabular}{crr}
      \topline
  \headcol 
    Parameter& Equation \eqref{eq::polynomial} & Equation \eqref{eq::polynomial::03}  \\
 \midline
\rowcol
       $a_0$   &$2.533317$& $-11.413655$\\
       $a_1$ & $1.960146$ & $12.323756$\\
 \rowcol   $a_2$&$2.118486$& $ -3.966212$\\
    $a_3$&$2.101649$&$0.454388$\\
    \hline
    \end{tabular}
\end{table}

\begin{table}[b!]
\centering
\caption{Bill of materials. Cost of the parts to assemble the low-cost sensors and electronics to measure plasma power and plasma brightness.}
{\small
 \begin{tabular}{ccc}
  \topline
  \headcol {Item} & Description & Unit cost (US\$) \\
\midline
R0   & R0-a and R0-b in parallel to obtain 52.8K$\Omega$   & --\\ 
\rowcol R0-a   & Resistor 56K$\Omega$ 
    RL07S563GRE5 (1/4 watt)  & 0.37\\ 
     R0-b& Resistor 931K$\Omega$ RN55D9313FB14 (1/8 watt) &  0.52\\ 
 \rowcol R1--R5 & Resistor 10M$\Omega$ VR68000001005JAC00  & 0.54\\ 
R6 & Resistor 2.2K$\Omega$ 1/4W 2\% & 0.30 \\ 
 \rowcol  R7--R12 & Resistor 10K$\Omega$ 1/4W 2\% & 0.30 \\
 R13 & Resistor 1K$\Omega$ 1/4W 2\% & 0.30 \\
 \rowcol  R14 & Resistor 3.3K$\Omega$ 1/4W 2\% & 0.30 \\
 C0   & Capacitor 1500pF 2kV 594-F152K53Y5RP63K7R   & 0.36 \\
    \rowcol  C1--C5  & Capacitor 15pF 3kV 
    564R30GAQ15 & 0.80 \\ 
    C6 & Electrolytic Capacitor 3$\mu$F 50V & 0.11\\
  \rowcol   C7 & Ceramic Capacitor 100pF & 0.34\\
     LDR & PDV-P9008 & 2.46\\
     \rowcol Shunt & Resistor 23$\Omega$ CCF6023R7FKE36 & 0.70\\
     TL072 & Operational amplifier &  0.69 \\
    \rowcol  TL431 & Precision Programmable Reference & 0.63 \\
     Battery & 9V Battery (two-set package) & 12.00\\
\rowcol Sensor & Light meter Urceri MT-912 & 30.00 \\
Needle & Tungsten Electrode 2\% Lanthanated & 10.00 \\
     \rowcol PCB & Solderable breadboard & 5.00\\
      Arduino & Arduino model Due & 40.00\\
\bottomlinec
\end{tabular}
}
 \label{tab::02}
\end{table}

\bibliography{alessandro}

\end{document}